\documentclass[preprint,showpacs,preprintnumbers,amsmath,amssymb]{revtex4}

\usepackage{color,graphicx}
\usepackage{dcolumn}
\usepackage{bm}

\begin{document}

\preprint{Preprint Copy}

\title{Microstructure, magneto-transport and magnetic properties of Gd-doped magnetron-sputtered amorphous carbon}

\author{L. Zeng}
 \email{lzeng@ucsd.edu}
  \affiliation{Materials Science and Engineering Program,University of California, San Diego, La Jolla, CA, 92093, USA}


\author{E. Helgren}
\author{F. Hellman}
 \affiliation{Department of Physics, University of California, Berkeley, Berkeley, CA, 94720, USA}
\author{R. Islam}
\author{D. J. Smith}
\affiliation{Center for Solid State Science and Department of Physics and Astronomy, Arizona State University, Tempe, AZ, 85287, USA}
\author{J. W. Ager III}
 \affiliation{Lawrence Berkeley National Laboratory, Berkeley, California, 94720, USA}


\date{\today}

\begin{abstract}
The magnetic rare earth element gadolinium (Gd) was doped into thin films of amorphous carbon (hydrogenated \textit{a}-C:H, or hydrogen-free \textit{a}-C) using magnetron co-sputtering. The Gd acted as a magnetic as well as an electrical dopant, resulting in an enormous negative magnetoresistance below a temperature ($T'$). Hydrogen was introduced to control the amorphous carbon bonding structure. High-resolution electron microscopy, ion-beam analysis and Raman spectroscopy were used to characterize the influence of Gd doping on the  \textit{a-}Gd$_x$C$_{1-x}$(:H$_y$) film morphology, composition, density and bonding. The films were largely amorphous and homogeneous up to $x$=22.0 at.\%. As the Gd doping increased, the $sp^{2}$-bonded carbon atoms evolved from carbon chains to 6-member graphitic rings. Incorporation of H opened up the graphitic rings and stabilized a $sp^{2}$-rich carbon-chain random network. The transport properties not only depended on Gd doping, but were also very sensitive to the $sp^{2}$ ordering. Magnetic properties, such as the spin-glass freezing temperature and susceptibility, scaled with the Gd concentration.
\end{abstract}

\maketitle

\section{\label{sec:level1}introduction}

Amorphous carbon (\textit{a-}C) thin films are amorphous semiconductors with a tunable band gap. Unlike other group IV amorphous semiconductors, such as \textit{a}-Si or \textit{a}-Ge, which have a tetrahedral $sp^{3}$-bonded random-network matrix, a typical \textit{a}-C film has a mixture of $sp^{2}$ and $sp^{3}$ bonding. Changing the $sp^2/sp^3$ ratio tunes the band gap of \textit{a}-C between that of diamond (100\% $sp^{3}$, band gap $E_{g}=5.4eV$) and graphite (100\% $sp^{2}$, semimetal with $E_{g}=0$). Studies of \textit{a}-C, diamond-like materials and carbon nanotubes have long focused on the superior mechanical properties. However, recent developments in novel electronic and spintronic materials have drawn attention to the electrical, optical and magneto-electronic properties~\cite{Schneider2004spintronics}. Room-temperature positive magnetoresistance (MR$\sim$60\%) was found in some magnetically-doped carbon thin films such as Co$_x$C$_{1-x}$~\cite{Xue2003CoC} and Ni$_x$C$_{1-x}$~\cite{Xue2003NiC}, with granular magnetic particles in an amorphous carbon matrix. Room-temperature positive MR was also reported for non-magnetically-doped (B-doped) polycrystalline diamond thin films (MR$\sim$100\%) ~\cite{Wang2000MRdia,Fei2002MRdia} and undoped \textit{a}-C/n-Si (MR$\sim$12\%) two-layer structures~\cite{Xue2005}. These results suggest potential technological applications of carbon-based thin films for electronics and spintronics applications. 

In this work, we have co-sputtered \textit{a}-C and the magnetic rare earth element Gd (\textit{a-}Gd$_x$C$_{1-x}$(:H$_y$)) with a wide range of \textit{x}, and studied the magnetic and magneto-transport properties. Gd has a half-filled \textit{f} electron shell and is trivalent in solids, which gives rise to a large local moment with \textit{J}=\textit{S}=7/2 and three valence electrons. Previous studies on Gd-doped e-beam co-evaporated amorphous Si have shown strong interactions between the magnetic moments and carriers leading to very large negative MR (e.g. 10$^{5}$ at 1 K), anomalous magneto-optical properties and spin-glass freezing [Refs \onlinecite{Teizer2000SSC, Helgren2005T*, Hellman1996PRL, Helgren2006scaling}]. The magnitude of this temperature-dependent negative MR below a characteristic temperature ($T^{*}$) was suggested to depend on electron screening related to the metallicity of the amorphous matrix~\cite{Helgren2005T*}. Screening is related to the band gap of the host semiconductor materials: narrower band gap corresponding to larger dielectric constant ($\epsilon$) causing more screening and reducing the effective interaction, leading to smaller MR and lower $T^{*}$. Evidence for this trend was that MR and $T^{*}$ were smaller for high $x$ and for \textit{a-}Gd-Ge compared to \textit{a}-Gd-Si samples~\cite{Sinenian2006GdGe}. These results suggest that the MR will be larger and occur at higher temperatures for a wider band gap matrix. One motivation of the present study was to make use of the tunable band gap of \textit{a}-C, which depends on the $sp^2/sp^3$ configuration, and then to study the magneto-transport properties of this type of thin film. 

The inclusion of Gd is shown to have large effects on transport and magnetic properties of the \textit{a}-C(:H) films. The dual roles of Gd in this material lead to strong interactions between the carriers and the magnetic moments, giving an enormous negative MR with strong temperature dependence at low temperature. Gd also has large influence on the carbon bonding thus film structure, which was studied using several complementary characterization methods. Correlations between the microstructure, bonding and other physical properties are discussed, and explanations for the properties are provided based on material and electronic structures.

\section{experimental details}

The \textit{a-}Gd-C(:H) films were prepared by co-sputtering high purity Gd and C targets onto substrates held at room temperature. Samples for ion-beam analysis were grown on MgO substrates, and samples for magnetization and magneto-transport measurements were grown on insulating amorphous SiN$_{x}$-covered Si substrates. Various Gd doping concentrations were achieved by controlling the ratio of the Gd/C flux. The base pressure prior to deposition was $\sim$5$\times$10$^{-8}$ Torr. The sputtering gas (Ar) working pressure was kept low (1.3$\sim$2.0 mTorr) in order to increase the mean free path of the sputtered atoms, thus preserving the initial kinetic energy of the sputtered target atoms, which is desirable for decreasing the $sp^{2}/sp^{3}$ ratio~\cite{Robertson2002report}. At this low sputtering pressure, the mean free path is comparable to the distance between cathode and substrate, so the number of collisions is $\sim$1. Therefore, the adatom energy is likely to be preserved near the initial energy (estimated to be 5-10 eV, depending on the sputtering power~\cite{Meyer1981}) gained from the Ar-target atom momentum-transfer process. It has been reported that the concentration of metastable $sp^{3}$ bonds increases with C atom energy and reaches a maximum in an energy window $\sim$100 eV~\cite{Robertson2002report}. Therefore, typical sputtered \textit{a}-C films are dominated by $sp^{2}$ sites. In order to suppress the graphitic $sp^{2}$ bonding and promote the $sp^{3}$ bonding, we introduced H atoms using a 95\%Ar:5\%H$_{2}$ mixture gas at the substrate in addition to pure Ar gas introduced at the Gd and C targets. Different H incorporation was achieved by changing the Ar/(95\%Ar:5\%H$_{2}$) flow ratio or by introducing a pure mixture 95\%Ar:5\%H$_{2}$ at the targets as the sputtering gas. 

Ion-beam analysis including Rutherford backscattering (RBS) with C resonance energy and hydrogen forward scattering (HFS) was used to determine the Gd, C and H concentrations. RBS also provided the atomic areal density from which atomic number density was obtained based on the film thickness as measured by Nanopics$^{\textregistered}$ atomic force microscope profilometer. High resolution cross-sectional transmission electron microscopy (HR-XTEM) with a JEOL JEM-4000EX was used to determine the film microstructure. Magnetization was measured in a Quantum Design MPMS$^{\textregistered}$ superconducting quantum interference device (SQUID) magnetometer. Magneto-transport measurements were carried out on samples with lithographically defined Hall bar patterns, with the magnetic field applied parallel to the film plane and to the applied current. This geometry leads to effects dominated by interactions with the Gd spin, rather than quantum backscattering.

\section{Material and Structural Characterization}

\subsection{High resolution electron microscopy}
\begin{figure}
    \centering
    \includegraphics[width=0.5\textwidth]{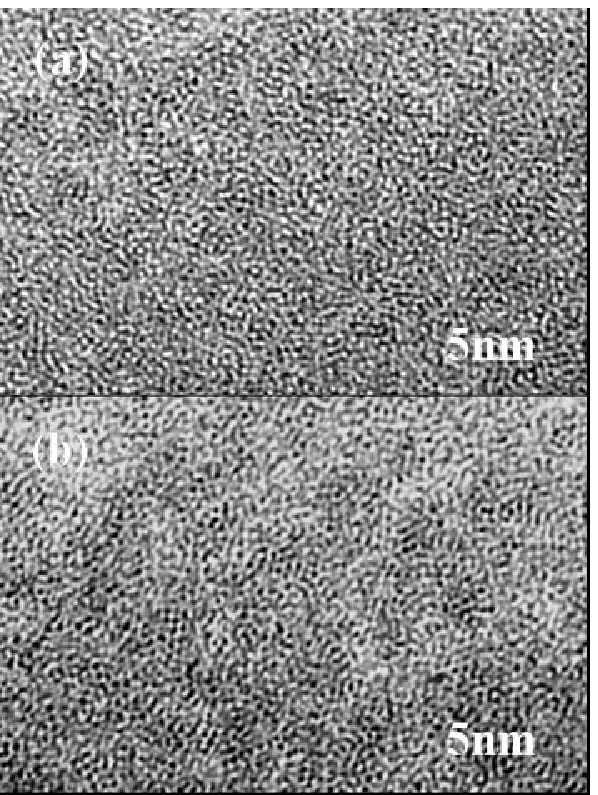}
            \caption{HR-XTEM micrographs for: (a) sputtered hydrogen free \textit{a}-C with 22.0 at.\% Gd (\textit{a}-Gd$_{0.22}$C$_{0.78}$); (b) sputtered hydrogenated \textit{a}-C:H with 15.6 at.\% Gd (\textit{a}-Gd$_{0.156}$C$_{0.844}$:H$_{0.20}$)}
            \label{fig:HRXTEM}
\end{figure}

Figure~\ref{fig:HRXTEM} shows typical HR-XTEM images for two samples. For both the hydrogen-free (Figure~\ref{fig:HRXTEM}(a)) and hydrogenated (Figure~\ref{fig:HRXTEM}(b)) Gd-doped films, the electron micrographs show dense amorphous microstructure, without any observable clustering or crystallization up to at least 22 at.\% Gd. The resulting films were very stable in ambient conditions. The same dense and featureless microstructure was also observed in other sputtered amorphous alloys prepared at low Ar pressure, such as \textit{a-}Gd-Si~\cite{Zeng2006PRB}. By contrast, \textit{a-}Gd-Si showed open columnar structure leading to massive oxidation when sputtered at Ar pressures of $\sim$6 mTorr. These results demonstrate that, under proper conditions, magnetron co-sputtering is an excellent method for obtaining a uniform distribution of dopants in amorphous group IV semiconductors (well above the thermodynamic equilibrium). 

\subsection{Rutherford backscattering and hydrogen forward scattering spectrometry}
\begin{figure}
    \centering
    \includegraphics[width=0.5\textwidth]{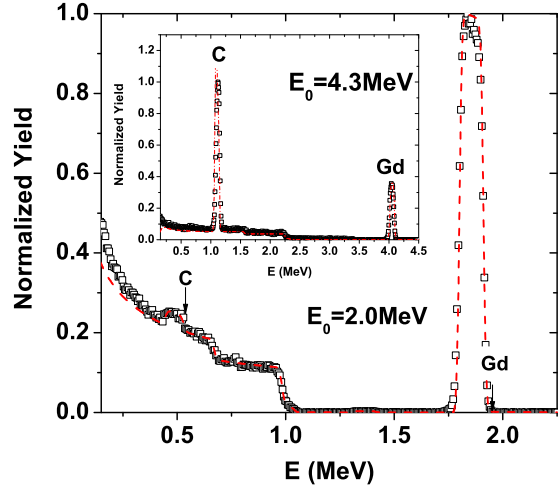}
            \caption{Typical RBS spectrum for Gd-doped H-free amorphous C film with incoming $^{4}$He energy $E_0$ at 2 MeV; Insert shows RBS spectrum for same sample with $E_{0}$ at carbon resonance energy of 4.3 MeV. The $\square$s are experimental data and dashed lines are simulation fits.}
            \label{fig:RBS}
\end{figure}

Figure~\ref{fig:RBS} shows a RBS spectrum for an \textit{a}-Gd-C sample. The backscattering yield $Y$ is proportional to $(Nt)$, the areal density of the element of interest, where $N$ is the number density (in atoms/cm$^{3}$) and $t$ is the film thickness. $Y$ is proportional to $Z^2$ so that low-$Z$ elements, such as carbon, have a much smaller yield than high-$Z$ elements, such as Gd. Moreover, the backscattering energy for light elements is low causing the signal to become convoluted with that from the substrate, as seen in the main plot of Fig.~\ref{fig:RBS}. The $^{4}$He energy was therefore tuned to 4.3 MeV, where $^{4}$He undergoes a nuclear resonance elastic scattering with C, thus increasing the backscattering cross-section of C and enhancing the C signal by a factor of $\sim$10, shown as the sharp peak in the Fig.~\ref{fig:RBS} insert. This technique measures the film composition with greater accuracy, and is essential to the quantitative study of any Gd concentration-dependent properties, such as the metal-insulator (M-I) transition or film magnetization. 

The H content in the \textit{a}-Gd-C:H films was measured by hydrogen forward scattering(HFS). The HFS samples were capped with a thin silver layer in order to separate the H signal in the sample from the signal of hydrocarbon contamination on the film surface. The hydrogen yields were calibrated with a Kapton sheet. For (95\%Ar:5\%H$_{2}$)/Ar gas flow ratio up to 10 \%, H incorporation was about 5 at.\% (\textit{a-}C$_{0.95}$:H$_{0.05}$) and was not very sensitive to the ratio. Films were also prepared with the pure mixture 95\%Ar:5\%H$_{2}$ gas introduced at the targets, which caused noticeable changes in the bonding structure indicated by Raman spectrum discussed below. The HFS experiment was not performed on these samples, but the H at.\% concentration was estimated to be high, $\sim$20 at.\% (\textit{a-}C$_{x}$:H$_{0.2x}$) based on Raman spectroscopy in the literature. Despite this high H concentration, the pure \textit{a}-C:H films were still very scratch resistant, as were the Gd-doped films which had a metallic appearance, showing no visible sign of polymerization. 

\subsection{Raman spectroscopy}

Raman spectroscopy is widely used to obtain information on $sp^{2}$ and $sp^{3}$ bonding in amorphous carbon films~\cite{Elman1982RamanC}. Disordered carbon systems have two Raman active modes when excited with a 488 nm excitation laser: one at 1580 cm$^{-1}$ of $E_{2g}$ symmetry, labelled `G'; the other at 1350 cm$^{-1}$ of $A_{1g}$ symmetry, labelled `D'. Both modes correspond to the $sp^{2}$ bonding, but are related to different atomic arrangement of the $sp^{2}$ sites. The G mode is the stretching vibration of any pair of $sp^{2}$-bonded sites, whether in chains or in rings, while the D mode is the collective breathing mode of the $sp^{2}$ sites in 6-member graphitic rings, thus only related to disordered graphitic sites. Depending on the $sp^{2}$ fraction and arrangement (whether in chains or in graphitic rings), the relative peak intensity, the peak position and the peak width of the two Raman modes will change accordingly. As $sp^{2}$ sites evolve towards $sp^{3}$ sites, the ratio of the D peak intensity to the G peak intensity ($I_{D}/I_{G}$) decreases monotonically~\cite{Robertson2002report}. When $sp^{3}$ sites start to nucleate, $I_{D}$ corresponding to graphitic bonds becomes negligible and the remaining $sp^{2}$ sites arrange mainly in chains.

\begin{figure}
    \centering
    \includegraphics[width=0.5\textwidth]{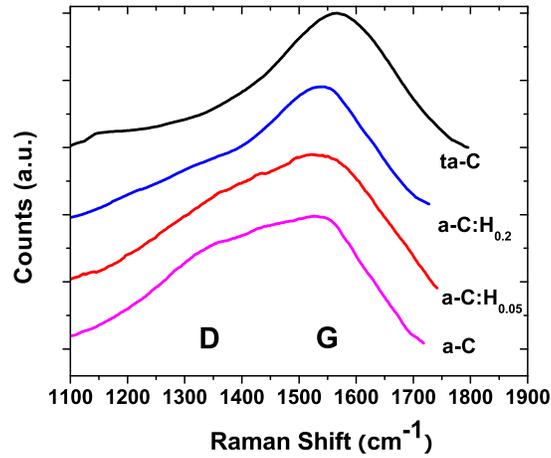}
            \caption{Raman spectra for different types of amorphous carbon. Note evolution of two Raman modes, D peak centered at $\sim$1390 cm$^{-1}$, and G peak centered at $\sim$1550 cm$^{-1}$. The \textit{ta-}C was prepared by cathodic-arc deposition, while the other films were prepared by magnetron sputtering.}
            \label{fig:RamanC}
\end{figure}

Figure~\ref{fig:RamanC} shows the Raman spectra for different types of amorphous carbon films, prepared by different growth methods (with and without H). We indeed observe two peaks, one centered at $\sim$1390 cm$^{-1}$ (the D peak) and the other centered at $\sim$1550 cm$^{-1}$ (the G peak). As the H concentration in the film is increased, the $I_{D}/I_{G}$ ratio decreases, indicating that H opens up the graphitic rings and helps to form $sp^{2}$-bonded carbon chains as expected from the literature. The width (full width half maximum from Gaussian peak fit) of the G peak, which scales with disorder, is 140-150 cm$^{-1}$ corresponding to a in-plane correlation length $L_{a}$ less than 0.1 nm~\cite{Robertson2002report}, further assuring the amorphous nature of the matrix. We compare our sputtered \textit{a}-C$_{1-y}$(:H$_{y}$) samples to a tetrahedral amorphous carbon film (denoted as \textit{ta}-C, with predominant $sp^{3}$-bonding) prepared by energetic vacuum cathodic-arc deposition (the top curve in Fig.~\ref{fig:RamanC}). A symmetric G peak with no sign of D peak indicates that there are no graphitic rings but only short $sp^{2}$-bonded chains remaining in a $sp^{3}$-dominant \textit{ta}-C matrix. Even though the $sp^{2}$ sites are a small fraction of the $sp^{3}$ sites in \textit{ta}-C, they still show a large G peak signal because the Raman cross-section for the $\pi$ bond is much larger than the $\sigma$ bond in $sp^{3}$. According to the three-stage model proposed by Ferrari~\cite{Ferrari2000RamanPRB}, the $sp^{2}$ fraction of our \textit{a}-C(:H) is about 90$\sim$100\%, typical for films prepared by sputtering. The incorporation of H changes the morphology of the matrix towards more ``diamond-like'' with less graphitic ring sites and $sp^{3}$ sites would have been stablized if a more energetic deposition method was used. 

\begin{figure}
    \centering
    \includegraphics[width=0.5\textwidth]{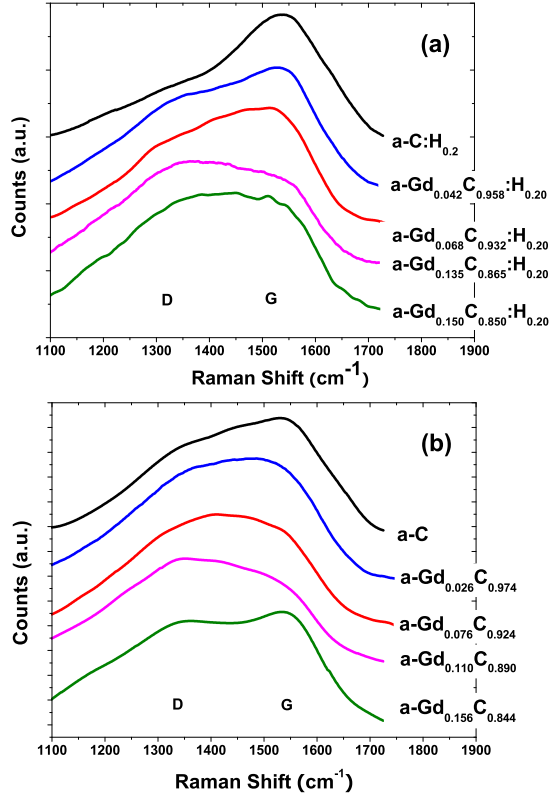}
            \caption{Raman spectra for sputtered Gd-doped (a) \textit{a}-C:H; (b) \textit{a}-C samples with different Gd doping. As Gd doping increases, $I_{D}/I_{G}$ increases first and then decreases. $I_{D}/I_{G}$ is larger than seen in any \textit{a}-C(:H) film in the literature.}
            \label{fig:RamanGdC3}
\end{figure}

Figure ~\ref{fig:RamanGdC3} shows the Gd influence on the \textit{a}-C(:H) structure. Although no changes were visible in the HR-XTEM micrographs as the Gd doping was increased (always amorphous and no clustering), there is a dramatic change of the $sp^{2}$ bonding order. As shown in Fig.~\ref{fig:RamanGdC3}, the $I_{D}/I_{G}$ ratio increases greatly with Gd concentration at first and then decreases. Starting as a low but broad shoulder next to the G peak in an undoped \textit{a}-C(:H), the D peak developes into a major Raman peak when Gd concentration increases. In contrast to the H incorporation which makes the \textit{a}-C more ``diamond-like'', Gd favors the formation of graphitic $sp^{2}$ sites. After the $I_{D}/I_{G}$ ratio reaches a maximum, it decreases slightly with the Gd concentration, but is still very large ($6\sim8$) compared to literature values which are $\leq$2.5~\cite{Robertson2002report}. The G peak width decreases after doping but is still broad (90-110 cm$^{-1}$, $L_{a}\sim$1 nm). These features were observed in both the hydrogenated (Fig.~\ref{fig:RamanGdC3}(a)) and H-free (Fig.~\ref{fig:RamanGdC3}(b)) samples. This structural change has a large consequence on the film transport properties as discussed later.

\section{Physical Properties}
\subsection{DC transport}

\begin{figure}
    \centering
    \includegraphics[width=0.5\textwidth]{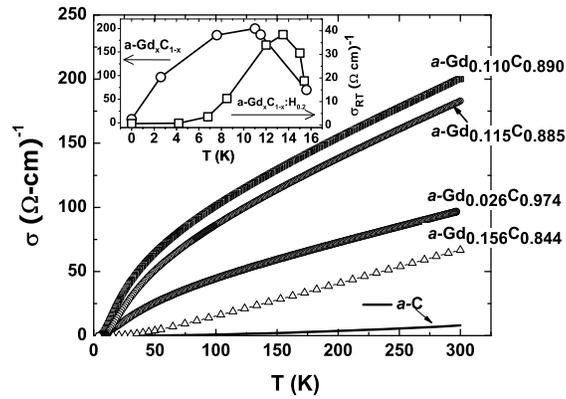}
            \caption{$\sigma_{DC}(T)$ of \textit{a}-Gd-C with different Gd doping. Inset shows $\sigma_{RT}$ as function of Gd at.\% for both H-free and hydrogenated samples. A peak in the $\sigma_{RT}$ is visible.}
            \label{fig:SigmaT2}
\end{figure}

The DC conductivity ($\sigma_{DC}$) and magneto-transport as a function of temperature ($T$) and magnetic field ($H$) was measured for the \textit{a}-Gd-C(:H) alloys. Figure ~\ref{fig:SigmaT2} shows that $\sigma_{DC}(T)$ for the pure H-free sputtered \textit{a-}C is very small, but greatly increases after Gd doping, thus the electrical doping effect of Gd is obvious. However, $\sigma_{DC}(T)$ increases with Gd doping only up to 11 at.\%Gd and then decreases with Gd doping. The 11 at.\% H-free sample which has the highest $\sigma_{DC}$ is very close to the M-I transition. \textit{a}-Gd$_{x}$C$_{1-x}$:H$_{y}$ behaves and scales similarly. The insert of Fig.~\ref{fig:SigmaT2} shows the room temperature conductivity ($\sigma_{RT}$) as a function of Gd at.\% for both H-free and hydrogenated samples. The only difference between them is that at similar $x$, \textit{a}-Gd$_{x}$C$_{1-x}$:H$_{0.2}$ has a much smaller $\sigma_{DC}$ due to the more insulating hydrogenated matrix. 

The high $T$ $\sigma_{DC}$ of \textit{a}-Gd$_{x}$C$_{1-x}$ has a linear slope and negative $d\rho/dT$ (as shown in Fig.~\ref{fig:SigmaT2}) as is found in other doped disordered semiconductors (Ref.~\onlinecite{Hertel1983PRLNbSi, 1997Yakimov_aMnSi, Liu2003PRBTbSi}). At low $T$, all samples show insulating behavior, with vanishing $\sigma_{DC}$ as $T\rightarrow$0 K. The transport mechanism is that of the variable-range-hopping (VRH) type: 
\begin{equation}
	\sigma=A exp\left[-\left(\frac{T_{0}}{T}\right)^{\nu}\right]
	\label{eq:VRH}
\end{equation}
where $\nu=1/4$ if electron-electron interactions are not taken into account (Mott VRH~\cite{Mott1968}), or $\nu=1/2$ if Coulomb repulsion between the carriers is included (Efros-Shklovskii VRH~\cite{Efros1975}). $\sigma_{DC}(T)$ at low $T$ fits $T^{1/2}$ better than $T^{1/4}$, similar to \textit{a-}Gd$_{x}$Si$_{1-x}$ alloys on the insulating side near the M-I transition~\cite{Teizer2000PRL, Xiong1999}.

\begin{figure}
    \centering
    \includegraphics[width=0.5\textwidth]{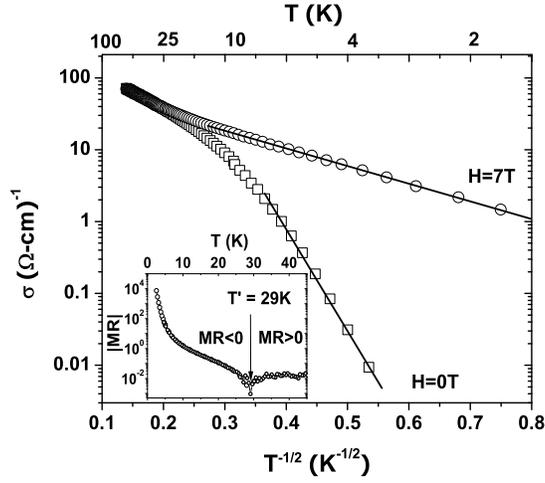}
            \caption{$\sigma_{DC}$ (on log scale) for \textit{a}-Gd$_{0.11}$C$_{0.89}$ vs. $T^{-1/2}$ in different $H$. Lines are fits to VRH with $\nu=-1/2$ at low $T$. Inset shows $\vert MR\vert = \vert\Delta\rho\vert/\rho = \frac{\vert\rho_{H}-\rho_{0}\vert}{\rho_{H}}$ (H=7 T) vs. $T$. Below $T'$=29K (indicated by the arrow), MR is large and negative; above $T'$, MR is small and positive.}
            \label{fig:MR1}
\end{figure}

Figure~\ref{fig:MR1} shows the semi-log plot of $\sigma_{DC}$ as a function of $T^{-1/2}$, measured in different $H$. An enormous increase of $\sigma_{DC}$ in 7 T ($\sim$10$^{4}$ at 2.5 K) can be seen. This enormous negative MR is found for both H-free and hydrogenated samples and behaves qualitatively the same. For brevity, results for a \textit{a}-Gd$_{0.11}$C$_{0.89}$ sample were shown to illustrate the two competing MR effects found in both types of films. One is a positive MR dominant at higher $T$ (up to 100 K). This is a small effect (MR value, defined as MR$= \Delta\rho/\rho = \frac{\rho_{H}-\rho_{0}}{\rho_{H}}$, is $\leq$0.02) and has very small $T$ dependence, as shown in the inset of Fig.~\ref{fig:MR1}. The other is a significant large negative MR, which increases very strongly as $T$ decreases. This negative MR dominates the small positive MR below a temperature $T'$ (=29 K, defined as MR=0 at 7 T for the \textit{a}-Gd$_{0.11}$C$_{0.89}$ sample). A similar $T'$ has also been found in Gd implanted \textit{ta}-C (\textit{ta}-C$_{1-x}$:Gd$_x$) samples~\cite{zeng2006MRS}. 

$T'$ plays the role of $T^{*}$ previously discussed for \textit{a}-Gd-Si~\cite{Helgren2005T*}, as the temperature below which Gd moments affect $\sigma_{DC}$, producing large negative MR~\cite{Helgren2005T*}. $T^{*}$ could not be defined for these (or other) insulating samples. $T'$ is the temperature below which the negative MR due to carrier-moment interactions dominates the positive MR which is due to electron correlation effect in disordered electronic systems~\cite{Lee1985}.  

\begin{figure}
    \centering
    \includegraphics[width=0.5\textwidth]{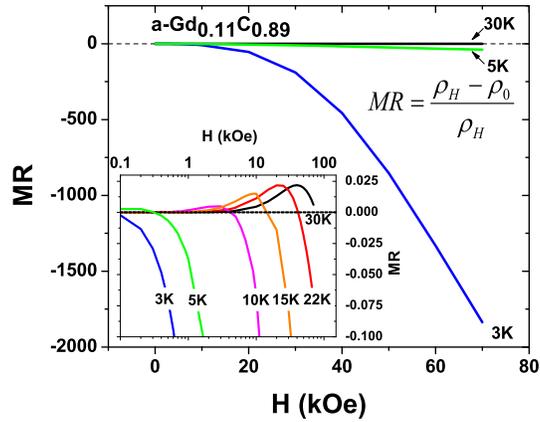}
            \caption{MR vs. $H$ on linear scale for $T$ = 3, 5 and 30 K respectively. Inset shows MR data on expanded semi-log scale near MR=0 for more temperatures. Small positive high $T$ MR and large negative low $T$ MR with a crossover between them at each temperature are visible.}
            \label{fig:MR2}
\end{figure}

Figure~\ref{fig:MR2} shows MR as a function of $H$ for the same sample. Large negative MR at 3 K, increasing with $H$, is visible. The inset expands the small MR region in order to show the positive MR as a function of the $H$ field. Three temperature regions can be identified. At high $T$ ($>$30 K), MR is positive up to 7 T. At moderate $T$ (5-22 K), MR is positive at low field and crosses over to negative MR at high field. The crossover field ($H_c$) decreases towards zero with decreasing $T$. Finally, at low $T$ (3 K), MR is strongly negative with a large $H$-dependence and does not saturate up to 7 T. 

\subsection{Magnetization}

\begin{figure}
    \centering
    \includegraphics[width=0.5\textwidth]{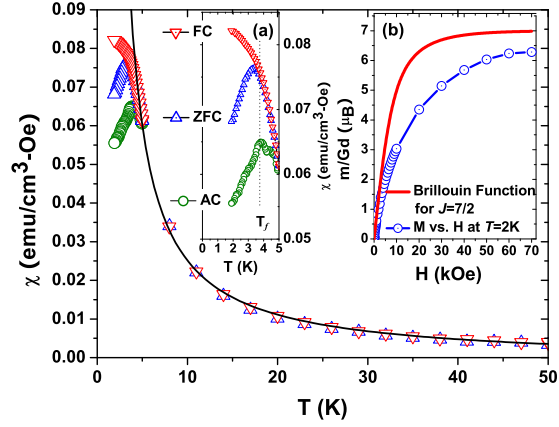}
            \caption{Magnetization data for sputtered \textit{a-}Gd$_{0.093}$C$_{0.807}$:H$_{0.05}$ sample. DC susceptibility $\chi=M/H$ for $H$=100 Oe measured after field cooling ($\chi_{FC}$) in 100 Oe and after zero field cooling ($\chi_{ZFC}$). AC susceptibility ($\chi_{AC}$) measured in 135 Hz 4 Oe. Curie-Weiss fit to data with $p_{eff}=$8.9 $\mu_{B}$ and $\theta=$1.2 K, shown as a solid line. Insert (a) shows expanded scale of low $T$ data ($T_{f}=$3.8 K). Insert (b) shows $M$ vs. $H$ at $T=$2 K for the same sample (line is a guide to the eye), compared to non-interacting Brillouin function.}
            \label{fig:magnetization}
\end{figure}

Figure~\ref{fig:magnetization} shows magnetization data for the \textit{a}-Gd$_{0.093}$C$_{0.807}$:H$_{0.05}$ sample. This data is representative of all \textit{a}-Gd$_{x}$C$_{1-x}$(:H$_{y}$) samples since they behave magnetically the same. A split was observed between susceptibility curves measured after field-cooled ($\chi_{FC}$) and after zero-field-cooled ($\chi_{ZFC}$), indicating spin-glass freezing, as previouly seen and extensively discussed in \textit{a}-Gd-Si alloys ~\cite{Hellman2000}. The spin-glass states are a result of competing ferromagnetic (FM) and antiferromagnetic (AFM) interactions in disordered materials. Since the \textit{f} shell of Gd is very tightly bound to the nuclei, direct interaction between the Gd moments is very unlikely. A RKKY (Ruderman-Kittel-Kasuya-Yosida) interaction was found in metallic Gd and GdSi$_{2}$ thus an indirect mediated RKKY-like exchange is likely the dominant interaction causing spin-glass freezing of our \textit{a}-Gd-C(:H) alloys. A freezing temperature $T_{f}$ (roughly where $\chi_{ZFC}$ and $\chi_{FC}$ data split) was confirmed by AC susceptibility ($\chi_{AC}$) measurements, as shown in Fig.~\ref{fig:magnetization} insert (a). $T_{f}$ and $\chi$ increase with Gd concentration, both in the \textit{a}-Gd-C samples and the \textit{a}-Gd-C:H samples. We note that all our samples are insulators according to the vanishing zero temperature conductivity, and are highly disordered (amorphous), making the nature of an RKKY interaction different from the conventional crystalline metallic materials. The localization length of electrons near the M-I transition is however significantly larger than the Gd-Gd spacing. 

Above $T_{f}$, in the paramagnetic state, $\chi_{ZFC}$ and $\chi_{FC}$ overlap and both fit well to the Curie-Weiss (CW) law:
\begin{equation}
	\chi_{ZFC,FC}=\frac{A}{T-\theta}.
	\label{eq:CW}
\end{equation}
where $\theta$ is the CW temperature and the constant $A=n_{Gd}p^{2}\mu_{B}^{2}/3k_{B}$. The effective moment $p$ (8.0 - 9.0 $\mu_{B}$) extracted from $A$ is slightly higher than the value for a free Gd$^{3+}$ ion (7.9 $\mu_{B}$) presumably due to polarization of carriers. There is no Gd concentration dependence of $p$. $\theta$ is very close to 0 K ($\left|\theta\right|\leq 4$ K), slightly increasing with Gd doping. H incorporation does not have any significant effect on $p$ or $\theta$. This is not like the \textit{a}-Gd$_{x}$Si$_{1-x}$ samples, where $p$ is suppressed and shows a nontrivial dependence on Gd concentration\cite{Hellman2000PRL}. No superparamagnetic or clustered glass behavior or enhanced effective moment was found, further evidence that Gd is uniformly distributed in the \textit{a}-C(:H) matrix and indirectly coupled via RKKY-like interaction. 

Fig.~\ref{fig:magnetization} insert (b) shows the $M$ vs. $H$ for the same sample. The high $H$ magnetization is below but close to the non-interacting Brillouin function for free Gd$^{3+}$ ion, confirming the AFM interactions in the film. However, the magnetization response to $H$ is much larger than the \textit{a}-Gd-Si samples, whose $M(H)$ curves are stongly suppressed below the Brillouin function due to much stronger AFM interactions~\cite{Hellman2000}. 

\section{Discussion}
As shown in the insert of Fig.~\ref{fig:SigmaT2}, $\sigma_{RT}$ increases then decreases, peaking at 11.0 and 13.5 at.\% for \textit{a}-Gd$_{x}$-C$_{1-x}$ and \textit{a}-Gd$_{x}$-C$_{1-x}$:H$_{0.2}$ respectively. The non-monotonic $x$ dependence of $\sigma_{RT}$ indicates that, in addition to electrically doping \textit{a}-C(:H), Gd has other effects on the film properties when the doping level is high, which cause the overall $\sigma_{DC}$ to decrease. As a result, a concentration-tuned M-I transition is not achieved for the \textit{a}-Gd$_{x}$C$_{1-x}$(:H$_{y}$) samples. In \textit{a}-Gd-Si alloys, $\sigma_{DC}$ decreased above 25 at. \% Gd, due to formation of nano-crystalline clusters observable in the HR-XTEM micrographs~\cite{Helgren2007}. Here, however, no crystalline clusters are observed in HR-XTEM micrographs (up to 22 at.\% and 15.6 at.\% for \textit{a}-Gd-C and for \textit{a}-Gd-C:H respectively). 

Another possible explanation of the $\sigma_{DC}$ drop is reduced electronic density due to low atomic number density ($n$) of the amorphous film. \textit{a}-C has many possible morphologies depending on the $sp^{2}/sp^{3}$ ratio and arrangement. Gd dopants could introduce large local defects (e.g. voids) and lead to less dense \textit{a}-C(:H) matrix, which should be reflected in a decrease of total number density ($n_{total}$) of the film. We however do not observe any significant drop of $n_{total}$ near the $\sigma_{DC}$ maximum. If we let six carbon atom replace one Gd atom in these doped films, we can estimate the effective mass density ($\rho$) of the \textit{a}-C(:H) matrix, which is between 1.9-2.6 g/cm$^3$, consistent with good quality sputtered \textit{a}-C(:H) films in the literature ~\cite{Robertson2002report}, and ruling out porous phase of \textit{a}-C(:H), which has very low electronic density thus very low $\sigma_{DC}$. 

\begin{figure}
    \centering
    \includegraphics[width=0.5\textwidth]{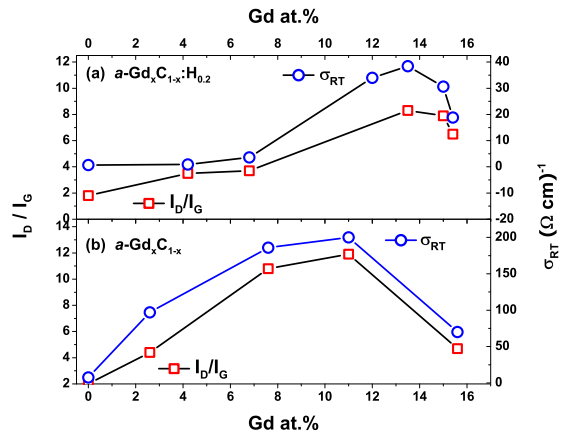}
            \caption{$\sigma_{RT}$ and $I_{D}/I_{G}$ as function of Gd concentration for (a) \textit{a}-Gd$_{x}$-C$_{1-x}$:H$_{0.20}$ (b) \textit{a}-Gd$_{x}$-C$_{1-x}$.}
            \label{fig:IDIG}
\end{figure}

We argue that the explanation for the non-monotonic behavior of $\sigma_{DC}$ can be deduced from the Raman spectra, which show an unambiguous change in the $sp^{2}$ arrangement with Gd doping, correlated with the concentration dependence of $\sigma_{RT}$ as shown in Figure~\ref{fig:IDIG}. The $I_{D}/I_{G}$ ratio first increases and then decreases, peaking at the same Gd concentration where $\sigma_{RT}$ reaches its maximum for both \textit{a}-Gd-C and \textit{a}-Gd-C:H. Samples with the highest $\sigma_{DC}$ have the largest $I_{D}/I_{G}$ ratio, indicating more disordered graphitic ring sites. These sites increase the $\sigma_{DC}$ of the \textit{a}-C(:H) matrix. The overall $\sigma_{DC}$ for the high Gd concentration samples can be treated as a sum of $\sigma$ of the \textit{a}-C(:H) matrix and of the Gd doping sites, as in a two-channel conductivity model. Before reaching the $\sigma_{DC}$ maximum, the increasing Gd at.\% and graphitic ring sites both contribute to the increasing $\sigma_{DC}$. However, the graphitic ring sites start to decrease after reaching the threshold doping and the matrix becomes less conducting. This compensates the doping effect of Gd and is responsible for the decreasing overall $\sigma_{DC}$.  

Turning now to the magneto-transport properties, there are few theoretical models to describe the large negative MR in magnetically-doped disordered electronic systems. The bound magnetic polaron (BMP) model was used to explain the negative MR in doped crystalline semiconductor systems such as Cd$_{1-x}$Mn$_{x}$Se~\cite{Wojtowicz1986} and Gd$_{3-x}\nu_{x}$S$_{4}$~\cite{Molnar1983PRL}. Our system however, has a much higher ratio of carriers to magnetic dopant sites compared to the above systems (of order 1:1 vs. 1:10$^{4}$). Thus, BMP is not a relevant model. Instead we turn to an Anderson localization type model to explain the negative MR. The interaction between carriers and the randomly oriented Gd moments provides additional magnetic disorder to the structural disorder thus increasing the localization of the hopping carriers. Applying $H$ reduces the magnetic disorder and increases the hopping conductance. This picture of magnetic disorder can qualitatively explain the large negative MR and emphasizes a single tunable parameter, namely the disorder, in these systems.

\section{Conclusion}
We have successfully prepared uniformly doped \textit{a}-Gd$_x$C$_{1-x}$(:H$_y$) samples with $x$ up to 22 at.\% and with variable amounts of H. For all samples, the magnetic ground state is a spin-glass with $T_{f}$ between 2-6 K, scaling with $x$. Above $T_{f}$, the magnetic susceptibility $\chi$ follows a Curie-Weiss law with $\theta\sim0$ K and $p_{eff}\sim$8.0 - 9.0 $\mu_{B}$, very close to the expected 7.9 $\mu_{B}$ for Gd$^{3+}$ and independent of $x$. $\sigma_{DC}(T)$ initially increases with $x$, but decreases above 11 at.\% (13.5 at.\% for the hydrogenated samples) and is strongly correlated to the amount of graphtic ring sites in films. Gd incorporation favors formation of graphitic $sp^{2}$ sites, while H incorporation opens up the graphitic rings, stablizes $sp^{2}$ chains and makes the \textit{a-}C(:H) matrix more ``diamond-like'' hence lower $\sigma_{DC}(T)$. Raman $I_{D}/I_{G}$ ratio after Gd doping is larger than most pure \textit{a}-C(:H) reported in the literature, suggesting that further modeling of the local bonding structure in this Gd-doped \textit{a}-C material would be appropriate. The large low-$T$, high-$H$ negative MR provides strong evidence of large interactions between the Gd moments and the charge carriers. A systematic study of the magnitude and onset temperature of the large negative MR by manipulating carbon matrix morphology towards more diamond-like (more $sp^{3}$ fraction, higher band gap) could shed light on the underlying physics and is currently under active investigation.

\begin{acknowledgments}
We thank K.M. Yu and B. Wilkens for useful discussions and help with the RBS data; D. Queen for assistance with the substrates. We acknowledge use of facilities in the John M. Cowley Center for High Resolution Electron Microscopy at Arizona State University. This research was supported by NSF DMR-0505524 and 0203907.
\end{acknowledgments}

\newpage 


\newpage

\end{document}